# Spin torque, tunnel-current spin polarization and magnetoresistance in MgO magnetic tunnel junctions

G. D. Fuchs[a], J. A. Katine[b], S. I. Kiselev[b], D. Mauri[b], K. S. Wooley[a], D. C. Ralph[a], and R. A. Buhrman[a]
[a]Cornell University, Ithaca NY 14853-2501
[b]Hitachi Global Storage Technologies, San Jose, California 95120

We examine the spin torque (ST) response of magnetic tunnel junctions (MTJs) with ultra-thin MgO tunnel barrier layers to investigate the relationship between the spin-transfer torque and the tunnel magnetoresistance (TMR) under finite bias. We find that the spin torque per unit current exerted on the free layer decreases by less than 10% over a bias range where the TMR decreases by over 40%. We examine the implications of this result for various spin-polarized tunneling models and find that it is consistent with magnetic-state-dependent effective tunnel decay lengths.

PACS: 72.25.–b, 75.75.+a, 85.75.-d

The ability of electron currents to transfer spin angular momentum, as well as charge, from one ferromagnetic electrode to another, and hence to exert a significant spin-torque on the electrodes (see e.g. refs. 1, 2), provides a powerful new tool for the study of spin transport in electronic structures, in addition to establishing new opportunities for future applications [2, 3]. The closely related issue of spin-dependent electron transport in magnetic multilayer structures, both magnetic tunnel junctions (MTJs) [4] and spin valves [5], is of wide-spread interest, both fundamentally and because the importance this phenomena has for information storage [6, 7]. A critical aspect of MTJs is the bias dependence of the tunnel magnetoresistance (TMR) [see e.g. ref. 8], which in general, decreases as the voltage bias ($V$) increases. Currently, there is no consensus as to a microscopic model that accounts for this behavior. Here we report our study of the relationship between bias dependent TMR and spin-transfer torque, which is fundamental to understanding both the nature of spin-polarized tunneling at finite bias and spin transfer effects in MTJs [9, 10, 11]. By making measurements of the thermally activated switching of nanostructured MTJs, we determine the bias dependence of the spin-torque transferred across an MgO tunnel barrier and its relation to the TMR. The spin-torque per unit current is, within 10%, a constant function of $V$ up to ±0.35 V in our devices, in contrast to the TMR which is reduced by >40% at ±0.35 V. This behavior is inconsistent with a decrease in the tunnel polarization factors of the electrodes as described by free-electron tunneling models [12, 13, 14], or by surface-magnon emission models that substantially decrease the surface magnetization with increasing bias [15]. We find, however, that magnetic-state-dependent tunneling decay lengths (effective masses) as theoretically predicted to result in very high TMR in MgO tunnel barriers [16, 17, 18], are consistent with our results, if we include the effects of our ultra-thin barrier layers having a high density of atomic defects and lower barrier heights than ideal MgO barriers.

We fabricated our MTJs by sputtering multilayer films on an unpatterned substrate that consisted of (in nm): Ta 3/Cu 28.5/Ta 3/PtMn 15.4/CoFe 1.9/Ru 0.7/CoFe 2.2/MgO 0.8/CoFe 1.0/Py 1.8/Ru 1.5/Ta 3.0/Cu 10, where CoFe is $Co_{90}Fe_{10}$ and Py is $Ni_{91.5}Fe_{8.5}$. The devices were formed into a nanopillar geometry using procedures described elsewhere [19]. The CoFe layer interfaced with PtMn is strongly exchange biased, as well as antiferromagnetically coupled to another CoFe layer through the Ru spacer via oscillatory exchange coupling [20], and therefore forms part of an exchange biased synthetic antiferromagnet (SAF). This SAF layer forms a nearly closed loop for the magnetostatic edge charges, thus producing only small (~0-140 Oe) dipolar fields $H_{dip}$ on the top, CoFe/Py, bilayer which forms the "free layer" for these devices.

Fig. 1 (a) shows the dc resistance TMR response of sample 1, which has a free layer in the form of a 50×100 nm elongated hexagon with an area $(A) = 3.5 \times 10^{-11}$ cm$^2$, as $H$ is swept through $H_{dip}$. The small magnetic volume of the free layer in this sample causes it to be thermally unstable at room temperature (RT), and so when $(H-H_{dip}) \sim 0$ there is almost no torque due to $H$, and the nanomagnet telegraphs between two stable states, nominally parallel (P, low resistance) and antiparallel (AP, high resistance) with respect to the reference layer. Throughout, we define $TMR = \Delta R / R_p$. Due to the low specific resistance of these junctions (~ 5 Ω–μm$^2$), the tunnel current at accessible bias is large enough to exert a spin-transfer torque $N_{st}$ on the free layer that is comparable in magnitude to the field torque $N_H$. $N_H$ and $N_{st}$ are not collinear, but nevertheless we can apply $H$ and $I$ so that $N_H$ and $N_{st}$ have equal but opposing effects on the energy barrier for magnetic reversal. This condition is met whenever the mean dwell times of the two stable resistance states are equal, which we determine from time traces of the telegraph behavior (Fig. 1 (b), inset). By varying $H$ we can examine the spin torque at bias levels ranging from very low $I$ up to the stability point of the tunnel barrier. In this range the TMR decreases from 33% (zero bias) to 19% (0.35 V bias).

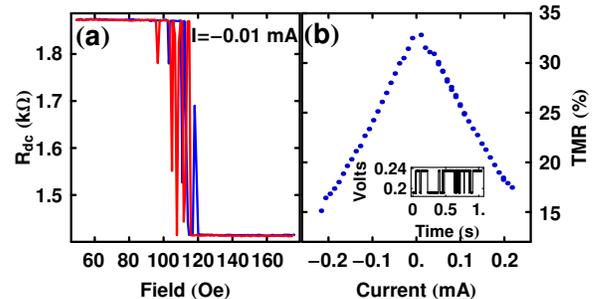

FIG. 1. (color online) (a) Plot of $R_{dc}$ vs. $H$ for sample 1 (b) Plot of TMR vs. $I$ for sample 1 calculated from telegraph traces (inset).

The RT equal dwell time $H$ vs. $I$ data for sample 1 is shown in Fig. 2 (a). The data follow a straight line for negative currents and positive currents up to $I \sim 0.1$ mA (region 1), after which there is a change in the behavior of the sample to a second linear region (region 2). In Fig. 2 (b) we show results for another device, sample 2, which also exhibits two regions of linear behavior with the break occurring in this device at $I \sim -0.03$ mA. Notice that here the break results in a higher linear slope for positive $I$, opposite to the case for sample 1. We made similar RT telegraph switching measurements on six other MTJ samples, and they all have similar behavior, typically exhibiting one break point in the linear equal-dwell-time behavior, with these breaks occurring at both positive and negative $I$.

Since the behavior of the equal-dwell-time data is central to this study we investigated the origin of the breaks in the data by cooling sample 1 to 5.6 K and found two distinct sets of nominally P and AP states that were occupied stochastically when $H$ was ramped to cycle the sample hysterically about its minor loop. These two sets of states had different coercive fields, as well as slightly different (<2% difference) resistance values for nominal P and AP alignment, indicating that the sample can have two slightly different micromagnetic states due to microcrystalline effects, and suggesting that the two regions in the $H$ vs. $I$ plot of equal dwell times also arise from two distinct micromagnetic states that form in response to different regions of $H$. Measurements of spin valve nanopillars with the same size and layer structure, with 2 nm of Cu replacing the MgO, also yielded kinks in $H$ vs. $I$ plots of equal dwell times, and measurements of the intrinsic anisotropy of the free layers in circular spin valve devices yielded intrinsic microcrystalline anisotropy fields that varied from 110 to 240 Oe in eight samples studied.

The relationship between the average dwell time of a Neel-Brown magnetic particle in a given magnetic state ($\tau_{p/ap}$) as a function of $H$ and $I$ is [21, 22, 23]:

$$\tau_{P/AP} = \tau_o Exp\left[\frac{E_a}{k_B T}\left(1 \pm \frac{H - H_{dip}}{H_{c,o}}\right)^\alpha \left(1 \mp \frac{I\gamma(I)}{I_{c,o}}\right)\right] \quad (1)$$

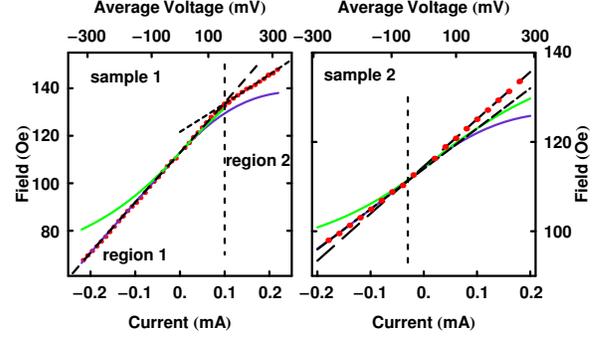

FIG. 2. (color online) Plot of values of $H$ and $I$ for equal dwell times of two level resistance fluctuations in sample 1 and 2. The top axis indicates the average of the voltage of the two states (tick spacing is not perfectly uniform due to a small non-linearity in the $I(V)$ characteristic). The diagonal dashed lines are fits with $\gamma(I)=1$ for region 1 and region 2. The solid purple or dark (green or light) lines are the calculated position of equal dwell times assuming $\gamma(I)$ is described by the free electron split band model (reduced surface magnetization model). The vertical dashed lines separate the two regions of linear behavior for each sample.

where, $\tau_o = 10^{-9}$ s is the attempt time, $E_a$ is the thermal activation barrier, $T$ is the temperature, $H_{c,o}$ is the $T=0$ coercive field, $I_{c,o}$ is the $T=0$ critical current, $\alpha$ is either 2 or 3/2 [24], and $\gamma(I)$ parameterizes how $N_{st}/I$ varies with $I$ (or with $V$). In addition, we normalize $\gamma(0)=1$. The condition $\tau_p = \tau_{ap}$ allows us to calculate the predicted $H(I)$ for equal dwell times, with the result being a linear relationship with a slope that depends on $I_{c,o}^{P \to AP}$, $I_{c,o}^{AP \to P}$, $H_{c,o}$ and $\alpha$, provided $\gamma(I)$=constant. Fig. 2 demonstrates that, apart from the breakpoints, the $H(I)$ data exhibit straight lines, indicating that the spin torque efficiency is indeed essentially constant over the bias ranges studied.

Fig. 3 (a) shows the equal dwell time plotted vs. $I$. Away from the micromagnetic break points, the dwell time data can be utilized to obtain values for the device spin torque parameters by fitting the data to Eq. (1). It is necessary to account for the significant self-heating in these low-specific-resistance MTJs, which varies as [25]: $T = \sqrt{T_{bath} + \beta I^2}$, where $T_{bath}$ is the bath temperature, and $\beta$ parameterizes the heating and is proportional to the resistance and geometry specific factors. For region 1 of sample 1, if we set $\gamma(I) = 1$ and assume $\alpha = 2$, we find that, $\beta = (8.5 \pm 0.3) \times 10^5$ K/mA$^2$, which is consistent with those found in previous MTJ ST studies, and $E_{a,1} = 0.51 \pm 0.01$ eV. This activation energy is comparable to, but as generally is the case, somewhat less than the nominal value of ~0.74 eV for ideal single domain rotation calculated by $E_a = H_{c,o} M_s Vol/2$, where $M_s$ is the saturation magnetization and $Vol$ is the volume of the free layer. We also find that for region 1, $I_{c,o}^{P \to AP} = 0.97 \pm 0.02$ mA, $I_{c,o}^{AP \to P} = 0.77 \pm 0.01$ mA, and $H_{c,o}$(region 1) = 350±10 Oe. This latter value is in very good agreement with the value $H_{c,o}$ (region 1) =348±10 Oe obtained by correcting the higher measured

Page 2

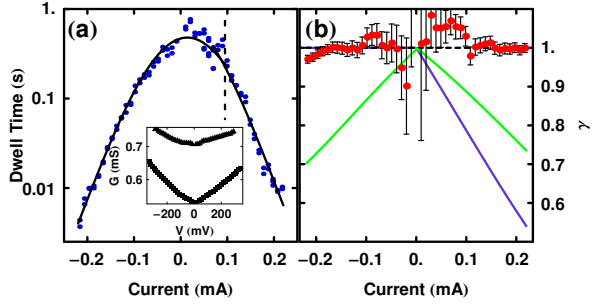

FIG. 3. (color online) (a) Dwell times of sample 1 plotted against *I*. Solid lines are fits in each of the two regions of magnetic behavior. The inset shows the conductance (*G=I/V*) of each state, calculated from the time traces, as a function of *V*. (b) $\gamma(I)$ data calculated as described in the text. The solid lines represents the prediction of $\gamma(I)$ calculated using the free electron split band model (purple or dark) and the reduced surface magnetization model (green or light) from the TMR(*I*) data.

T=5.6 K coercive field for sample 1 for the change in saturation magnetization ($M_s$) between 5.6 K and room temperature (RT) [26]. Assuming that the critical currents and $\beta$ are the same in region 2 as in region 1, we obtain $E_a$ = 0.53±0.01 eV and $H_{c,o}$ = 285±10 Oe, where the latter is also in excellent agreement with the other coercive field determined from the low *T* measurement, $H_{c,0}$ = 282±10 Oe.

To further test our conclusion that $\gamma(I) \sim 1$ and $N_{st}/I$ ~constant within our experimental error over our accessible range of applied bias ($V \sim \pm 0.35$ V), we extract $\gamma(I)$ explicitly from the data (Fig. 3(b)). This $\gamma(I)$ data is determined by subtracting the $H(I)$ fit from the $H(I)$ data, which is then scaled into $\gamma(I)$ through Eq. (1). At small *I*, the uncertainty is larger since both *I* and $(H-H_{dip})$ are small, and therefore deviations due to uncertainty in the determination of *H* at equal dwell times becomes more important. As Fig. 3(b) indicates, the $\gamma(I)$ we calculate from the data is either a constant function of *I* or decreasing by less than 10%.

The constancy of $\gamma(I)$ over the accessible bias range (0 – 0.35 V) has implications for various models used to interpret the decrease in TMR(*V*) observed in MTJs. In split-band free electron models [12, 13] the decrease in TMR with increasing *V* is due to changes in the effective polarization ($P_{col}$) of the collector electrode at energy eV above the Fermi level. The effective polarization of the emitter ($P_{emit}$) in these models remains constant, as the electrons tunnel from the same set of states near the Fermi energy. Since the TMR, given by Julliere's formula [4], $TMR = \Delta R/R_p = 2P_{col}P_{emit}/(1-P_{col}P_{emit})$, is sensitive to the product of the polarization factors, it decreases for both directions of *I*. In contrast, in these models the magnitude of $N_{st}/I$ (and hence $\gamma(I)$), on a given layer is proportional to only the effective polarization of the counter electrode [11]. Since we designed our devices so that only the free layer is responsive to spin-torque, there is an asymmetry imposed by *V*. For negative *I*, the reference layer is the emitter, whose polarization factor remains constant. For positive *I*, however, the reference layer is the collector, which according to the free electron model has a decreasing polarization factor with increasing *V* (Fig. 2 (a), inset). Using Julliere's formula, the free electron model's prediction for $\gamma(I)$ can be calculated from the TMR data [27]:

$$\gamma(I) = \frac{TMR(I)}{2+TMR(I)}\left(\frac{2+TMR_{max}}{TMR_{max}}\right) \text{ for } I > 0; \gamma(I)=1 \text{ for } I < 0. \quad (2)$$

This predicted form of $\gamma(I)$, calculated from our TMR data, is plotted in Fig. 3 (b) while Fig. 2 shows the prediction for the $H(I)$ equal dwell time data. The prediction that $\gamma(I)$ should be reduced to ~ 0.6 at $I \sim 0.2$ mA, is clearly inconsistent with our measurements.

A second proposed mechanism for TMR(*V*) is hot electron emission of surface magnons that either reduces the surface magnetization through the population of surface magnon modes [15], or results in spin-flip during the tunneling process [28]. If the first mechanism is a symmetric process, it will decrease the polarization factors of the electrodes by an equal amount, and therefore:

$$\gamma(I) = \sqrt{\frac{TMR(I)}{2+TMR(I)}\left(\frac{2+TMR_{max}}{TMR_{max}}\right)}, \quad (3)$$

which, at highest bias, would cause $\gamma(I)$ to decrease to ~0.7 (Figure 3 (b)). The effect of this form of $\gamma(I)$ on the $H(I)$ data for sample 2 is plotted in Fig. 2 (b), which shows a substantial deviation from the measurement result. If the surface magnons are only populated predominately on one of the two electrodes there would be an even greater deviation from a linear $H(I)$, in one bias direction or the other.

The spin-torque that might be exerted on the free layer electrode by electrons undergoing spin flip via magnon emission during tunneling cannot be calculated within the formalisms that have been published to date [28]. Until such a calculation is undertaken we can only comment that it seems unclear how such a process would maintain the same transverse spin torque provided by electrons that only tunnel without spin flip. Similarly, the effects of impurity assisted resonant tunneling [29] do not simultaneously account for the conduction behavior as well as for a constant $\gamma(I)$ using reasonable resonance line-widths in our devices, although resonant tunneling may be playing some role in these devices.

The very large TMR that has been achieved with epitaxial Fe/MgO/Fe MTJ and related systems [30, 31] is attributed to symmetry considerations in the overlap of the spin-dependent electronic wave functions between the Fe electrodes and MgO tunnel barriers [16, 17, 18]. In bcc Fe, majority electrons have either s-like or pd-like character. When the electrodes are aligned in the P state, the s-like electrons tunnel through the MgO with a smaller effective mass $m^*$, *i.e.* a longer tunnel decay length, than those with pd-like character, and as a result, it is the s-like electrons that dominate the majority-to-majority tunneling processes in thick, well-formed barrier layers. The minority electrons, having pd-like character, have a much lower tunneling rate, yielding a high tunnel polarization. When the electrodes are aligned in AP configuration, in the assumed case of coherent tunneling, i.e. in the absence of scattering, only the majority (minority) electrons with pd-like character can tunnel into minority (majority) states of the collector with the same symmetry. Due to the band structure of the MgO, $m^*$ for these electrons is higher and hence the tunnel decay length is shorter. This gives rise to two related effects; the predicted TMR is very high due to the reduced conductance in the AP configuration, and the different $m^*$'s result in the AP conductance increasing more rapidly with $V$ at high bias than the P conductance.

The ultra-thin-barrier MgO MTJ's studied here have far from ideal TMR behavior with ~30% TMR at low bias, indicating the coherent tunneling process has been significantly diluted. Recent scanning tunneling spectroscopy studies have found that textured (001) ultra-thin MgO layers grown by sputtering or e-beam evaporation on Fe and CoFe (001) surfaces have a substantial density of oxygen vacancies, with a higher density in the thinner layers, presumably due to an un-relaxed lattice mismatch. In ~ 1 nm layers, these vacancies result in conduction and valence band tails extending to ±0.5 eV of the Fermi level [32]. The presence of such defect states introduces the possibility of momentum scattering processes such as co-tunneling, which allow some pd-like electrons to scatter elastically and tunnel as s-like states and vice-versa, reducing the TMR. In addition, since our bias range (±0.35 V) is a significant fraction of the tunnel barrier height (~0.5 eV), the high $m^*$ electrons, both majority and minority, that tunnel coherently and dominate in the AP configuration only, have a strong voltage-dependent conductance increase, causing a strong bias dependence of the TMR. Since such momentum scattering is spin independent, these effects enhance the tunnel conductance without affecting the spin current, which leads to an approximately constant $N_{st}/I$ for each magnetic configuration that is consistent with our experimental result. Fitting the $I(V)$ characteristic of each magnetic state with the Simmons model [33] yields a significantly larger $m^*$ for the AP configuration, and is consistent with this picture of the conduction.

In conclusion, we have studied the bias-dependence of the spin-torque response of a MTJ with an ultra-thin MgO tunnel barrier and a thermally unstable free layer at room temperature. The spin-torque transferred between the reference layer and the free layer per unit current decreases less than 10% up to ~ ± 0.35 V, at which point the TMR level has decreased by over 40%. This observation sheds new light on the process of TMR reduction with bias by imposing additional constraints on any model that describes this process. We find that the data can be well described by a magnetic-orientation-dependent difference between the effective masses for majority tunneling, which causes the conductance to increase more rapidly in the AP state than in the P state, while preserving the polarization of the tunnel current in each state, and hence the efficiency of the spin-transfer torque at increasing bias.

We thank P. Brouwer and Shaffique Adam for very useful conservations and P. Mather for sharing the STS data. This work is supported by ARO/MURI, DAA19-01-0-0541, and by NSF through the NSEC support of the Center for Nanoscale Systems. Additional support provided by NSF through use of the NNIN/CNF facilities and the facilities of the Cornell MRSEC.